# Layer-dependent interlayer antiferromagnetic spin reorientation in air-stable semiconductor CrSBr

Chen Ye[1,2,†], Cong Wang[3,†], Qiong Wu[4,†], Sheng Liu[5], Jiayuan Zhou[6], Guopeng Wang[6], Aljoscha Söll[7], Zdenek Sofer[7], Ming Yue[4], Xue Liu[1,*], Mingliang Tian[6], Qihua Xiong[8,9,10,11,*], Wei Ji[3,*], Xiao Renshaw Wang[2,12,*]

[1]*Key Laboratory of Structure and Functional Regulation of Hybrid Materials of Ministry of Education, Institutes of Physical Science and Information Technology, Anhui University, Hefei 230601, China*
[2]*Division of Physics and Applied Physics, School of Physical and Mathematical Sciences, Nanyang Technological University, Singapore 637371*
[3]*Beijing Key Laboratory of Optoelectronic Functional Materials and Micro-Nano Devices, Department of Physics, Renmin University of China, Beijing 100872, China*
[4]*Faculty of Materials and Manufacturing, Key Laboratory of Advanced Functional Materials, Ministry of Education of China, Beijing University of Technology, Beijing 100124, China*
[5]*Okinawa Institute of Science and Technology, Okinawa Prefecture 904-0412, Japan*
[6]*School of Physics and Optoelectronics Engineering, Anhui University, Hefei 230601, China*
[7]*Department of Inorganic Chemistry, Faculty of Chemical Technology, University of Chemistry and Technology Prague, Technická 5, 16628 Prague 6, Czech Republic*
[8]*State Key Laboratory of Low-Dimensional Quantum Physics and Department of Physics, Tsinghua University, Beijing 100084, P.R. China*
[9]*Frontier Science Center for Quantum Information, Beijing 100084, P.R. China*
[10]*Beijing Academy of Quantum Information Sciences, Beijing 100193, P.R. China*
[11]*Beijing Innovation Center for Future Chips, Tsinghua University, Beijing 100084, P.R. China*
[12]*School of Electrical and Electronic Engineering, Nanyang Technological University, 50 Nanyang Ave, 639798, Singapore*

[†]These authors contributed equally
[*]Email: lxue@ahu.edu.cn; qihua_xiong@tsinghua.edu.cn; wji@ruc.edu.cn; renshaw@ntu.edu.sg

# Abstract

Magnetic van der Waals (vdW) materials possess versatile spin configurations stabilized in reduced dimensions. One magnetic order is the interlayer antiferromagnetism in A-type vdW antiferromagnet, which may be effectively modified by the magnetic field, stacking order and thickness scaling. However, atomically revealing the interlayer spin orientation in the vdW antiferromagnet is highly challenging, because most of the material candidates exhibit an insulating ground state or instability in ambient conditions. Here, we report the layer-dependent interlayer antiferromagnetic spin reorientation in air-stable semiconductor CrSBr using magnetotransport characterization and first-principles calculations. We reveal an odd-even layer effect of interlayer spin reorientation, which originates from the competitions among interlayer exchange, magnetic anisotropy energy and extra Zeeman energy of uncompensated magnetization. Furthermore, we quantitatively constructed the layer-dependent magnetic phase diagram with the help of a linear-chain model. Our work uncovers the layer-dependent interlayer antiferromagnetic spin reorientation engineered by magnetic field in the air-stable semiconductor.

**Keywords:** antiferromagnetic semiconductor, CrSBr, magnetoresistance, interlayer reorientation, layer-dependent

# Introduction

Magnetic van der Waals (vdW) materials, which integrate layered vdW materials with long-range magnetic orders, have stimulated a surge of interest because of their magnetic interactions.[1–4] In reduced dimensions, versatile arrangements of preferred spin orientations[5–7] are stabilized by the interplay between exchange interaction and magnetic anisotropy. And the spin orientation can be effectively engineered by different physical modulations, including magnetic field, electric field, temperature and thickness. Among various types of magnetic orders, the interlayer antiferromagnetism, in which spins are aligned as ferromagnetic (FM) order in each layer and antiferromagnetic (AFM) order across the vdW gap, is fundamental yet technologically advantageous. First, the interlayer AFM exchange coupling generates a more stable and faster dynamics of spin texture than that in the ferromagnet.[8] Second, such an antiparallel interlayer orientation leads to the parity-time symmetry preserved in even-layered antiferromagnet but broken when layer number is odd. Consequentially, the symmetry breaking can give rise to various fascinating physical phenomena as well as be engineered by a wide variety of approaches.[3,9–11] Third, the reduction of dimensionality in vdW antiferromagnet enables emergent quantum phenomena, including layer Hall effect[9] and giant nonreciprocal second-harmonic generation.[12] Despite the merits, most reported vdW antiferromagnets show either insulating in transport properties or air instability,[4,13,14] hindering the investigation of layer-dependent interlayer AFM spin reorientation.

CrSBr, which is an A-type antiferromagnet with a relatively high Néel temperature ($T_N$ ~ 140 K) preserved down to bilayer thickness, attracts great attention owing to its exotic properties, such as magnetic field-controlled excitonic effects, negative magnetoresistance and large magnetic proximity effect.[15–20] In addition, two outstanding features of CrSBr, namely air stability and semiconductivity, position it as a proper candidate of material to reveal and manipulate the layer-dependent interlayer coupling. To be specific, magnetic vdW materials are prone to degrade in the air through oxidation or hydration,[21,22] which evidently obstacles the accurate characterizations. Thus, an air-stable vdW antiferromagnet is required for disclosing the intrinsic layer dependency.[23] Furthermore, CrSBr shows a semiconducting behaviour with a desirable conductivity at low temperature, allowing various approaches to characterizing and manipulating the magnetic properties. The rare combination of air-stable magnetism and semiconductor couples the spin and charge degrees of freedom,[24–26] allowing electrically probing spin configuration.

However, when downscaling the thickness to an atomic level, the interlayer AFM coupling between CrSBr layer remains unknown.

In this work, we report the layer-dependent interlayer spin reorientation in air-stable semiconductor CrSBr using magnetotransport characterizations and an AFM linear-chain model. The negative magnetoresistance establishes the link between interlayer spin orientation and electronic properties in CrSBr. Furthermore, an odd-even layer effect in the spin-flop transition was observed, which is ascribed to the energy difference from the uncompensated magnetization in odd-number CrSBr thin flakes. Moreover, we captured the magnetic phase diagram reflecting the interlayer spin reorientation under a magnetic field through the linear-chain model.

## Results

The CrSBr single crystals were synthesized using a chemical vapour transport method (in Supplement Information). Figure 1a shows the schematic of the experimental setup. We mechanically exfoliated the atomically thin CrSBr from bulk crystal and stamped it onto a silicon wafer covered with a 285 nm thick $SiO_2$. The Cr/Au electrodes were deposited for the following magnetotransport measurements. Figure 1b shows the layered crystal structure of CrSBr. Each buckled CrS plane is sandwiched by the top and bottom Br atoms and stacks along the *c*-axis *via* vdW interaction. The rectangular structure from the *c*-axis perspective benefits the identifying of crystal axes morphologically. The layer number *N* of CrSBr was precisely determined by an optical contrast based on the pixel red-green-blue value. Figure 1c presents the optical micrograph of CrSBr multilayers with *N* from 1 to 5. Figure 1d shows the optical contrast map of the same region of exfoliated CrSBr with a 631 nm optical filter. The optical contrast linearly scaled with thickness from 17.5 for monolayer to 61.5 for pentalayer (red dots in Fig. 1e) and the corresponding thickness was determined by the atomic force microscopy (blue triangles, Fig. 1e). Figure 1f presents the highly preserved crystallinity of exfoliated multilayers characterized by transmission electron microscopy (TEM). Notably, the CrSBr multilayers were prepared and stored in an ambient environment over one month without degradation confirmed by time-dependent Raman spectra and optical contrast (Figs. S3-S6), supporting the excellent air stability of CrSBr. Moreover, we performed the Raman spectroscopy as an additional approach to identify the $N$. Figure 1g shows the Raman spectra of CrSBr multilayers as a function of $N$ at room temperature. The two peaks, namely $P_1$ and $P_2$ at around 247.5 cm$^{-1}$ and 346.2 cm$^{-1}$, respectively, were retained down to a monolayer thickness. A layer dependence was also clearly observed that $P_1$ and $P_2$ downshifted by ~10 and ~2 cm$^{-1}$ with a reducing $N$, respectively.

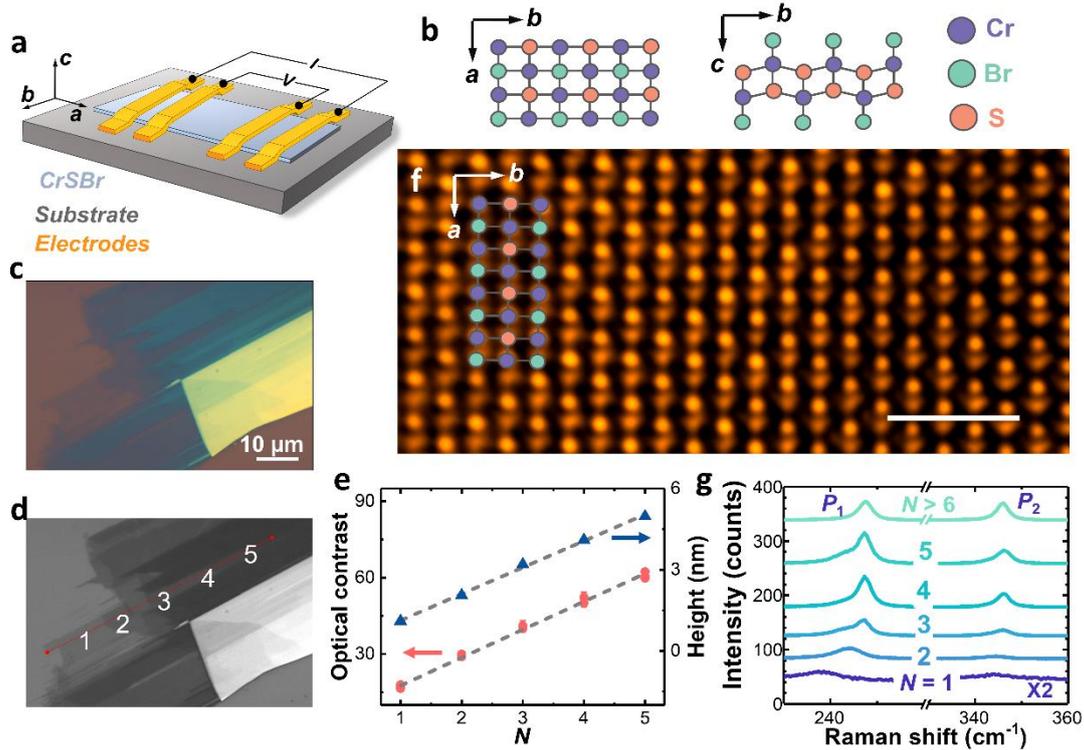

Figure 1. Identification of layer thickness and structural characterizations. (a) Schematic of the magnetotransport measurements on the CrSBr device. (b) Crystal structure of CrSBr. The left image shows a top view from the $c$-axis and the right image shows a side view from the $a$-axis. Purple, green and orange balls represent the Cr, S and Br atoms, respectively. (c) Optical image of CrSBr multilayers. The scale bar is 10 μm. (d) Optical contrast map of the same CrSBr multilayers with a 631 nm optical filter.

(e) Extracted optical contrast with different layer numbers *N* of the CrSBr multilayers. (f) Transmission electron microscopy (TEM) image of CrSBr multilayers along the *c*-axis at room temperature. The scale bar is 1 nm. (g) Layer-dependent Raman spectra of CrSBr multilayers. The position of the two peaks shows a layer dependence that downshifts gradually with a reducing *N*. The spectra intensity of monolayer CrSBr is multiplied by a factor of two for clarity.

To probe the interlayer spin orientation in the CrSBr, we conducted magnetotransport characterizations as functions of magnetic field and temperatures. The spin orientation correlating with the interlayer tunnelling determines the behaviour of resistance (*R*) under the magnetic field,[27,28] making magnetoresistance a probe to capture the interlayer AFM dynamics. Figures 2a, 2b and 2c show the temperature-dependent magnetoresistance of CrSBr (*N*=4) in magnetic fields along the *a*-, *b*- and *c*-axis, respectively. The orange and black dashed lines denote the critical ($H_c$) and saturation fields ($H_s$) of the spin-flop transition, respectively. The negative magnetoresistance reveals that interlayer tunnelling coupled with spin orientation transited from suppression to restoration. At a sufficient low magnetic field, the spins are forced to be anti-aligned along the easy axis, leading to the suppression of interlayer tunnelling. With an increasing magnetic field, the spin-flop transition took place with a slight canting of spins restored the suppressed interlayer tunnelling gradually. Afterwards, the saturation is achieved as fully-polarized spins point along the direction of the magnetic field.[15,29] Therefore, the restoration of interlayer tunnelling became saturated.

Figures 2d, 2e and 2f show the colourmaps of temperature-dependent magnetoresistance ratio (MRR), which can be expressed as $MRR = 100\% * [R(H) - R(0)]/R(0)$, extracted from Fig. 2a, 2b and 2c, respectively. The clear edges of the dome denote the $H_s$. Besides, we noted that a clear plateau, indicated by the orange dashed line in Fig. 2e, emerged when the magnetic field along the *b*-axis. Figure S13 shows the colourmap of MRR retrace (magnetic field from positive to negative), in which a plateau also appeared.

This feature reveals that a finite energy is required to trigger the spin-flop transition along *b*-axis in the even-*N* CrSBr without a global magnetization. This feature of spin-flop transition is different from the weak anisotropy scenario.[4] In the weak anisotropy case, $H_c$ is almost zero as observed in even-*N* $CrCl_3$.[4] Therefore, the observed nonzero $H_c$ in CrSBr suggests an uncovered energy competition in the 2D vdW antiferromagnet.

Figure 2g shows the temperature-dependent *R* of CrSBr (*N* = 4) from 2 to 280 K. The *R* increases as the temperature decreases, confirming the semiconducting transport property of CrSBr. The temperature-dependent *R* is well described by the thermal activation model at high temperatures (in Fig. S14). And a kink of derivative of *R* at *T* ~ 135 K (in Fig. S15) is consistent with the MRR results, determining $T_N$ ~ 135 K of CrSBr flake (*N* = 4). Figure 2h shows the temperature-dependent $H_s$ extracted from the black dashed lines in Fig. 2d-f. It is evident that $H_s$ decreases as the temperature increases, and $H_s$ becomes negligible when above $T_N$ ~ 135 K, because the thermal fluctuation tends to misalign the collinear spin orientation at high temperatures. The unequal $H_s$ in three axes throughout the varied temperatures confirms the triaxial anisotropy of CrSBr. The easy, intermediate and hard axis of the CrSBr is along *b*-, *a*- and *c*-direction, respectively. Figure 2i shows the temperature-dependent MRR (*H* = $H_s$). The MRR shows a nonmonotonic temperature dependence with a kink at near 40 K, below which the MRR upshifted suddenly. Such a peculiar feature was observed in similar CrSBr works[15,30] and it is ascribed to the magnetic defects.[30]

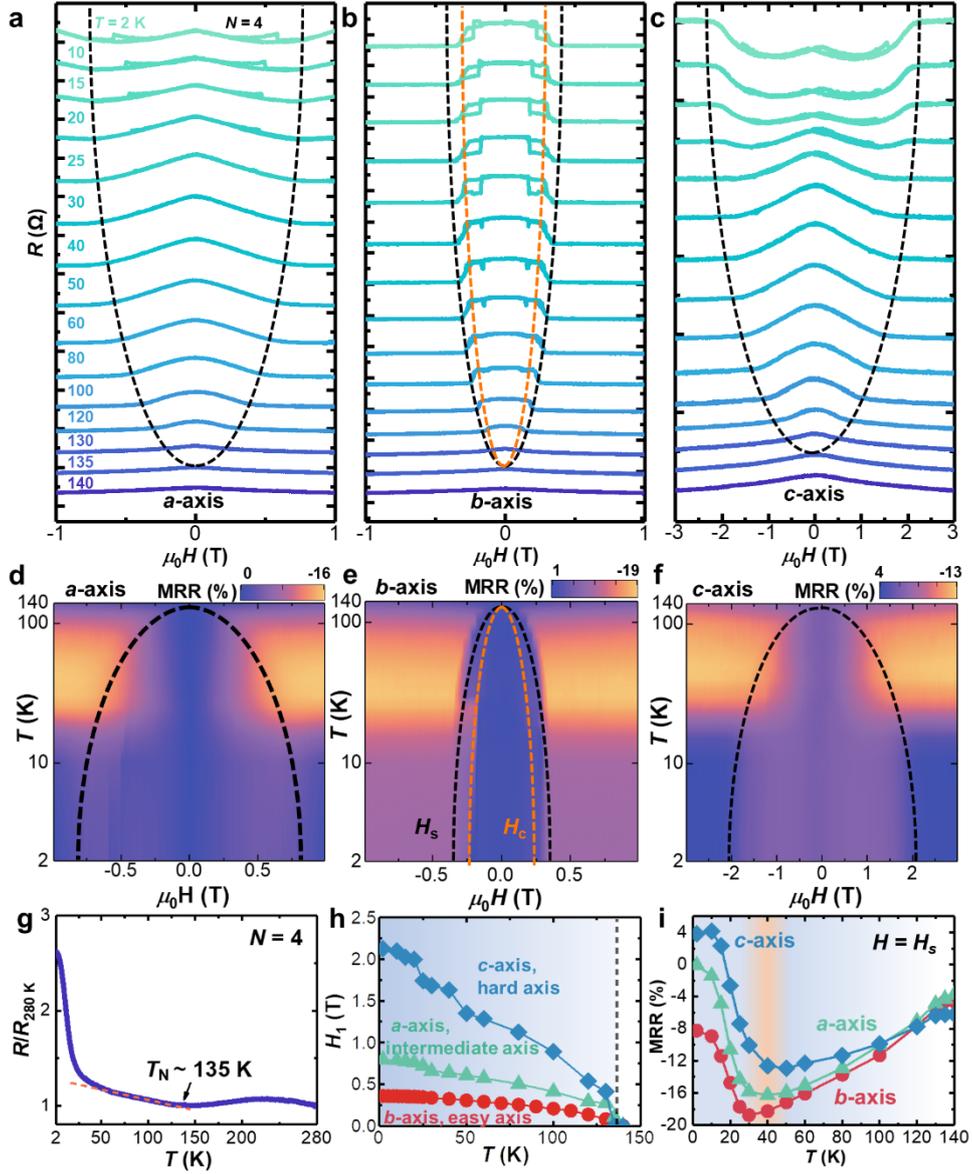

Figure 2. Temperature-dependent magnetotransport measurements of tetralayer CrSBr. Magnetoresistance as a function of temperature and magnetic fields along *a*-axis (Fig. 2a), *b*-axis (Fig.2b) and *c*-axis (Fig. 2c). The trace and retrace of magnetoresistance are presented and offset for clarity. Scale bar in Fig. 2a-c is 0.1 MΩ. The black dashed lines denote the temperature-dependent saturation field ($H_s$), and the orange dashed line (Fig. 2b) denotes the critical field of the spin-flop transition ($H_c$). The colourmap for the trace of magnetoresistance ratio (MRR) as a function of temperature under magnetic fields along *a*-axis (Fig. 2d), *b*-axis (Fig.2e) and *c*-axis (Fig. 2f). The black and orange dashed lines denote the $H_s$ and $H_c$, the same as indicated in Fig. 2a-c. (g) Temperature-dependent normalized resistance ($R/R_{280\ K}$). The kink at 135 K indicates the Néel temperature ($T_N$) for AFM phase transition. (h) Temperature-dependent $H_s$. The different magnitudes from the *a*-, *b*- and *c*-axis confirm the triaxial anisotropy of CrSBr. (i) Temperature-dependent magnetoresistance ratio extracted for $H = H_s$. A peculiar kink shows the maximum at 40 K as indicated by the orange region.

To reveal the layer-dependent AFM spin reorientation of CrSBr, we collected magnetoresistance with thickness scaling from $N = 2$ to 6. We also applied the AFM linear-chain model[4,31] to interpret the interlayer spin reorientation. The linear-chain model treats the magnetization of each layer as one macro spin and the spins couple between adjacent layers through interlayer exchange coupling. Therefore, the spin orientation determined by internal energy competition can be written as

$$E = \mu_0 M_S \left[ \frac{H_J}{2} \sum_{n=1}^{N-1} cos(\varphi_n + \varphi_{n+1}) - \frac{H_K}{2} \sum_{n=1}^{N-1} (cos\varphi_n)^2 - H \sum_{n=1}^{N-1} cos\varphi_n \right], \quad (1)$$

where $M_S = 2g\mu_B S$ denotes the saturated magnetization of a single layer, and $\varphi_n$ is the angle of macro spin in the *n*-th layer. The $H_J = 2J/(\mu_0 M_S)$ and $H_K = K/(\mu_0 M_S)$ are the exchange energy and magnetic anisotropy energy in the magnetic field scale, respectively. Figure 3a shows the magnetoresistance as a function of *N* measured at 2 K. The magnetic field was applied along the *b*-axis. The dark and light colours for each *N* denote the trace and retrace of magnetoresistance, respectively. The spin-flop transitions were observed with a clear hysteresis loop in all thicknesses. In odd-*N* CrSBr multilayers, it is easy to understand that the difference originates from the uncompensated magnetization. For *N* is even, the hysteresis loops may result from two different spin states but with the same energy configuration[31] in the trace and retrace scans, or defects-induced domain wall pinning effect[32]. Furthermore, figures S3-S6 show the temperature dependency for each thickness ($N = 2, 3, 5, 6$), confirming that the interlayer antiferromagnetism is retained down to bilayer thickness. The corresponding $T_N$ was plotted in Fig. 3b, showing independence to *N* as reported.[16]

Moreover, figure 3c shows the prominent odd-even layer effect, which manifests as the $H_c$ oscillation against the increasing *N*. It confirms that spin-flop transition in odd-*N* multilayers with a higher critical energy than that in even-

*N* multilayers. This parity-dependent $H_c$ shows consistency to our proposal that additional Zeeman contribution of an uncompensated layer requires a larger $H_c$ for the odd-*N* multilayers. We also noted that $H_c$ generally increases with *N*, besides the oscillation feature. Likewise, Fig. 3d shows that $H_s$ increases as the *N* increases. The $H_s$ for the fully-polarized FM state can be described in the AFM linear chain model, which is expressed as

$$H_s = 2H_J cos^2\left(\frac{\pi}{2N}\right) - H_K. \qquad (2)$$

Hence, the $H_s$ depending on the $H_J$ and $H_K$ is expected to increase drastically when *N* decreases, and it approaches the limit $2H_J$-$H_K$ when *N* is sufficiently large. Our experimental data shows a drastic reduction of $H_s$ when below a tetralayer thickness, corresponding to a smaller energy required to evolve AFM state to FM state. This is ascribed to the layer-dependent *J* decreased clearly below a tetralayer thickness (Table S1), which might be a signal of dimensional crossover from 3D to 2D. Figure 3d shows the experimental $H_s$ (diamond dots) extracted from Fig.3a as a function of *N* at 2 K. The grey line denotes theoretical fitting. The $H_s$'s trend shows consistency with the theoretical prediction of the linear-chain model, from which the $H_J$ and $H_K$ were estimated to be 0.266 T and 0.099 T, respectively. Hence, the $H_s$ of bulk form is estimated to 0.433 T, which is closed to the $H_s$ = 0.47 T as reported.[15] The consistency further validates the AFM linear-chain model and affirms the AFM spin reorientation in the CrSBr. Figure 3e exhibits the layer-dependent difference of $H_s$ ($\Delta H_s$), which is $\Delta H_s = H_s^c - H_s^b$, obtained from magnetoresistance under the *b*-and *c*-axis magnetic field at 2 K. The AFM linear-chain model predicts that

$$\Delta H_s = H_{ext} - \gamma M_V, \qquad (3)$$

where $H_{ext}$ is the internal effective field for the fully polarized spins. The γ and $M_V$ denote the demagnetization factor and magnetization per unit cell volume, respectively. For atomically thin devices, γ=0 under the in-plane magnetic field and γ=1 for an out-of-plane magnetic field.[4] Therefore, the constant $\Delta H_s$ implicates a constant $H_{ext}$ irrespective of the magnetic field direction. This isotropic $H_{ext}$ reveals that the shape anisotropy from magnetic dipole-dipole interaction plays a crucial role of determining the interlayer antiferromagnetism in CrSBr,[33] which is different from other 2D magnets dominated by the magnetocrystalline anisotropy.[13,34,35]

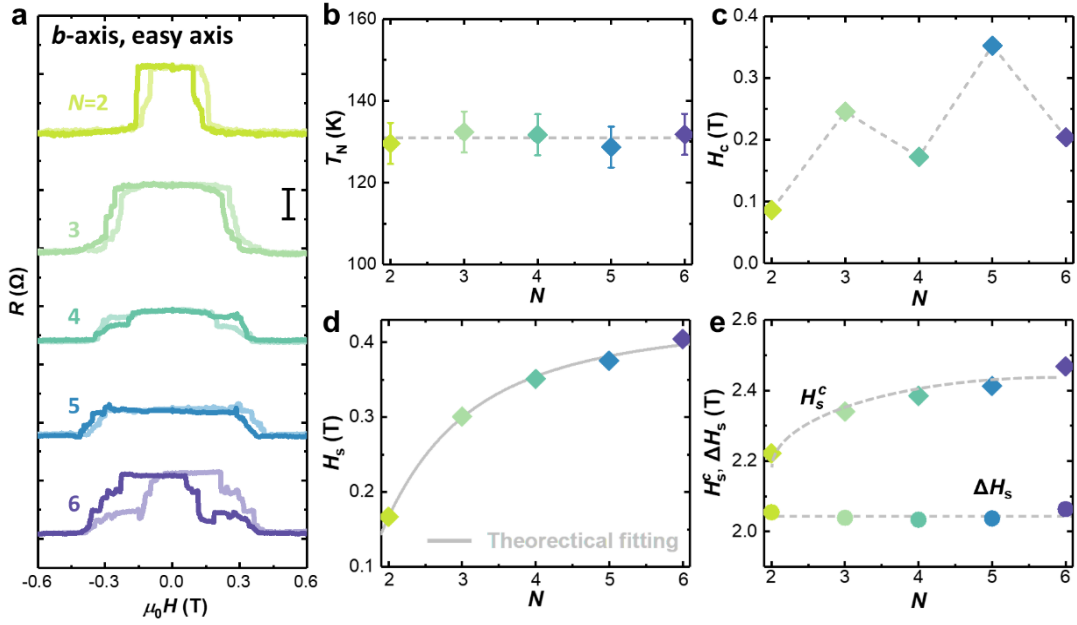

Figure 3. Layer-dependent spin-flop transitions of CrSBr multilayers. (a) Layer-dependent magnetoresistance along the *b*-axis at 2 K. The light and dark colours in each scan denote the trace and retrace of magnetoresistance, respectively. The scale bar is 0.05 MΩ. (b) Néel temperature ($T_N$) of CrSBr with different thicknesses. (c) Thickness-dependent $H_c$ at 2 K extracted from Fig. 3a. An odd-even effect was observed due to the uncompensated magnetization of one layer in the odd-*N* CrSBr multilayers. (d) Thickness-dependent $H_s$ at 2 K extracted from Fig. 3a. The data were well fitted by the antiferromagnetic linear-chain model to obtain the magnetic anisotropy and interlayer exchange coupling in the magnetic field scale. (e) Thickness-dependent $H_s$ under the *c*-axis magnetic field at 2 K. The difference of $H_s$ between two magnetic field directions $\Delta H_s$ shows the independence of $H_s$ against *N*.

Density-functional theory (DFT) calculations were carried out to reveal the origin of the layer-dependent interlayer spin reorientation under the *b*-axis magnetic field. The monolayer FM groundstate, fewlayer A-type AFM groundstate and the easy axis along the *b*-axis direction were verified in our calculations (Figs. S16, S17, Tables T1, T2). Layer-dependent behaviors were depicted for the nearest interlayer spin-exchange parameter (*J*) and the magnetic anisotropy energy ($K$) in Fig. 4a and 4b, respectively. While $J$ and $K$ oscillates with different behaviors at odd-even layer numbers. We thus infer that the competitions among $J$, $K$ and the extra Zeeman energy of uncompensated magnetization in odd layers result in the experimentally observed layer-dependent magnetic properties. Furthermore, Monte Carlo simulations were performed using the AFM linear-chain model with inputs from our DFT results (Fig. 4a and 4b) to predict the evolution of spin orientation. Layer-dependent $H_{c\text{-DFT}}$ and $H_{s\text{-DFT}}$ of spin-flop transitions were summarized in Fig. 4c and the trend is consistent with the transport observation. Figure 4d shows the magnetic phase diagrams of tetralayer CrSBr under the *b*-axis magnetic field. When the magnetic field reaches the $H_c$, the collinear AFM orientation (Fig. 4e) abruptly reorients to a noncollinear alignment (Fig. 4f) which gradually cants under further increasing fields and eventually forms the FM state (Fig. 4g) at $H_s$. The arrows denote the macro spins in individual layer, and $\varphi_n$ is the rotated angle for the *n*-th layer. Notably, the noncollinear spin orientation are different between odd- and even-*N* CrSBr. For even-*N* multilayers (Fig. 4d-g), the noncollinear orientation arrange symmetrically after the spin-flop transition (Fig. 4f), *i.e.* $\varphi_1 = \varphi_4$ and $\varphi_2 = \varphi_3$. Whereas the asymmetric noncollinear spin orientation occurs when *N* is odd (Fig. 4j), *i.e.* $\varphi_{1,5} \neq \varphi_{2,4}$, owing to the uncompensated magnetization. Layer-dependent evolution of interlayer spin reorientation with *N* ranging from 2 to 6 are calculated (Fig. S18).

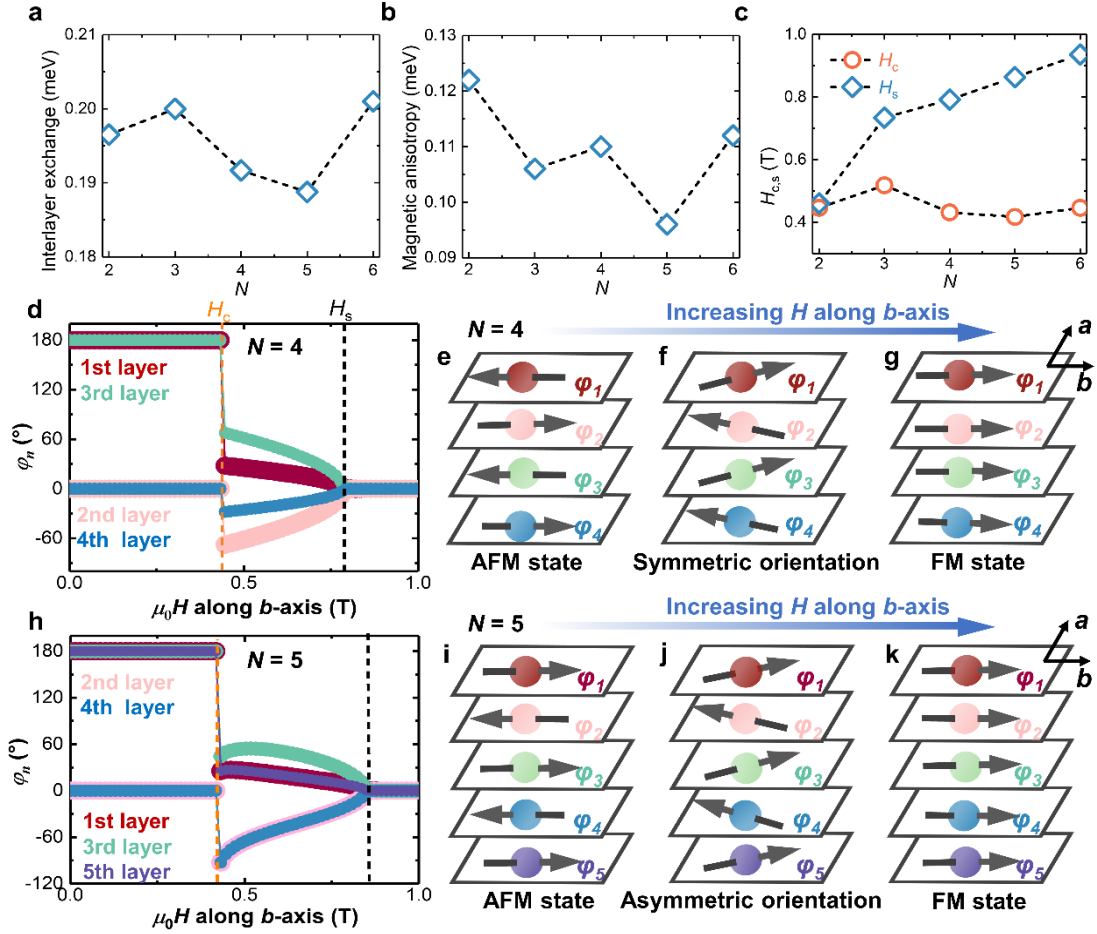

Figure 4. Interlayer spin reorientation of CrSBr multilayers. (a) Layer-dependent interlayer exchange energy. (b) Evolution of magnetic anisotropy energy as a function of layer number $N$. (c) Theoretical predicted layer-dependent critical field ($H_{c\text{-DFT}}$) and saturation field ($H_{s\text{-DFT}}$) of spin-flop transitions. (d,h) The magnetic phase diagrams of tetralayer ($N = 4$, Fig. 4d) and pentalayer ($N = 5$, Fig. 4h) are calculated using the antiferromagnetic linear-chain model. The phase diagram reflects the interlayer spin reorientation under the $b$-axis magnetic field. (e,f,g) Schematic of spin reorientation in a tetralayer CrSBr from an AFM state to the noncollinear intermediate state and then to the FM state. (i, j, k) Schematic of interlayer spin reorientation in a tetralayer CrSBr from an AFM state to the noncollinear intermediate state and then to the FM state. Due to the uncompensated magnetization, the noncollinear spin orientation shows an asymmetric orientation in the odd-$N$ CrSBr multilayers.

## Conclusions

In summary, we tracked the interlayer AFM spin reorientation in the air-stable semiconductor CrSBr down to bilayer thickness. Using magnetotransport characterization and an AFM linear-chain model, we provided a quantitative and thorough diagram of the interlayer spin reorientation as functions of layer thickness and external magnetic field. The spin-flop transition occurs in all thicknesses and directions of the magnetic field. We revealed the odd-even layer effect of the spin-flop transition field stemming from the competition of interlayer exchange, magnetic anisotropy energy and Zeeman energy. Furthermore, through the linear-chain model, we quantitatively determined the layer-dependent magnetic phase diagram of CrSBr. Our work provides insights into interlayer antiferromagnetism of CrSBr on the parameter space of layer number, temperature and magnetic field.

## Methods

### Bulk crystal synthesis

The CrSBr single crystals were synthesized through a chemical vapour transport (CVT) method. Chromium, sulfur and bromine in a stoichiometric ratio of CrSBr were added and sealed in a quartz tube under a high vacuum. The tube was then placed into a two-zone tube furnace. The pre-reaction was done at 700 °C for 10 hours. Then, the source and growth ends were kept at 850 and 900 °C, respectively. After 25 hours, the temperature gradient was reversed, and the hot end gradually increased from 880 to 950 °C over 5 days. High-quality CrSBr single crystals with lengths up to 2 cm were achieved.

### Device fabrications

Followed by standard processes of electron beam lithography and thermal evaporation, we have fabricated CrSBr devices with a typical four-probe geometry for the magnetoresistance measurements. Because CrSBr flakes usually form a rectangular bar shape after exfoliation, we have carefully aligned the current probes either along the crystalline a- or b-axes. The contact metals used for all devices are 4 nm Cr and then 30 nm Au. Electrical contacts were made by ultrasonically bonding Al wires onto the electrodes on the devices.

**Electrical characterizations**

Transport measurements were conducted in an Oxford TeslatronPT cryostat with a temperature range from 300 to 2 K and a maximum magnetic field of 8 T. Keithley 2400 sourcemetre and 2000 multimetre were employed for the magnetotransport measurements.

## Supplementary Information

Supplementary information available: More structural characterization of CrSBr ultrathin flakes, magneto-transport characterization of CrSBr with layer number *N* from 2 to 6, DFT calculations and details of linear-chain model.

## Acknowledgment

X.R.W. acknowledges support from Academic Research Fund Tier 2 (Grant No. MOE-T2EP50120-0006 and MOE-T2EP50220-0005) and Tier 3 (Grant No. MOE2018-T3-1-002) from the Singapore Ministry of Education, and the Agency for Science, Technology and Research (A*STAR) under its AME IRG grant (Project No. A20E5c0094). Z.S. was supported by project LTAUSA19034 from Ministry of Education Youth and Sports (MEYS). A.S. received funding from the European Union's Horizon 2020 research and innovation programme under grant agreement No. 956813. X.L. acknowledges support from the Natural Science Foundation of China (12104006) and the Natural Science Foundation of Tianjin (19JCQNJC00700). Q.X. gratefully acknowledges the funding support from State Key Laboratory of Low-Dimensional Quantum Physics and Tsinghua University start-up grant. W.J gratefully acknowledges financial support from the Ministry of Science and Technology (MOST) of China (Grant No. 2018YFE0202700), the National Natural Science Foundation of China (Grants No. 11974422), the Strategic Priority Research Program of Chinese Academy of Sciences (Grant No. XDB30000000), the Fundamental Research Funds for the Central Universities, China, and the Research Funds of Renmin University of China [Grants No. 22XNKJ30]. C.W was supported by the National Natural Science Foundation of China (Grants No. 12104504) and the China Postdoctoral Science Foundation (2021M693479). Calculations were performed at the Physics Lab of High-Performance Computing of Renmin University of China, Shanghai Supercomputer Center.

## Author contribution

C.Y. performed the electrical measurements with the help of X.L.. A.S. and Z.S. synthesized the bulk crystal of CrSBr. J.Z., S.L., J.Z., G.W. and X.L. fabricated the devices and characterized the CrSBr multilayers. Q.W. and M.Y. performed the STEM characterizations. C.W. and W.J. provided the theoretical calculation. C.Y., C.W. and X.R.W draft the paper and revise the paper with the help of X.L., M.T., Q.X. and W.J.. All authors contributed to the scientific discussion. X.R.W designed and directed the study.

## References

(1) Burch, K. S.; Mandrus, D.; Park, J.-G. Magnetism in Two-Dimensional van Der Waals Materials. *Nature* **2018**, *563* (7729), 47–52.

(2) Fei, Z.; Huang, B.; Malinowski, P.; Wang, W.; Song, T.; Sanchez, J.; Yao, W.; Xiao, D.; Zhu, X.; May, A. F.; Wu, W.; Cobden, D. H.; Chu, J.-H.; Xu, X. Two-Dimensional Itinerant Ferromagnetism in Atomically Thin $Fe_3GeTe_2$. *Nat. Mater.* **2018**, *17* (9), 778–782.

(3) Zhong, D.; Seyler, K. L.; Linpeng, X.; Wilson, N. P.; Taniguchi, T.; Watanabe, K.; McGuire, M. A.; Fu, K.-M. C.; Xiao, D.; Yao, W.; Xu, X. Layer-Resolved Magnetic Proximity Effect in van Der Waals Heterostructures. *Nat. Nanotechnol.* **2020**, *15* (3), 187–191.

(4) Wang, Z.; Gibertini, M.; Dumcenco, D.; Taniguchi, T.; Watanabe, K.;

Giannini, E.; Morpurgo, A. F. Determining the Phase Diagram of Atomically Thin Layered Antiferromagnet CrCl$_3$. *Nat. Nanotechnol.* **2019**, *14* (12), 1116–1122.

(5) Yu, X. Z.; Koshibae, W.; Tokunaga, Y.; Shibata, K.; Taguchi, Y.; Nagaosa, N.; Tokura, Y. Transformation between Meron and Skyrmion Topological Spin Textures in a Chiral Magnet. *Nature* **2018**, *564* (7734), 95–98.

(6) Lu, X.; Fei, R.; Zhu, L.; Yang, L. Meron-like Topological Spin Defects in Monolayer CrCl$_3$. *Nat. Commun.* **2020**, *11* (1), 4724.

(7) Gomonay, O.; Baltz, V.; Brataas, A.; Tserkovnyak, Y. Antiferromagnetic Spin Textures and Dynamics. *Nat. Phys.* **2018**, *14* (3), 213–216.

(8) Gibertini, M.; Koperski, M.; Morpurgo, A. F.; Novoselov, K. S. Magnetic 2D Materials and Heterostructures. *Nat. Nanotechnol.* **2019**, *14* (5), 408–419.

(9) Gao, A.; Liu, Y.-F.; Hu, C.; Qiu, J.-X.; Tzschaschel, C.; Ghosh, B.; Ho, S.-C.; Bérubé, D.; Chen, R.; Sun, H.; Zhang, Z.; Zhang, X.-Y.; Wang, Y.-X.; Wang, N.; Huang, Z.; Felser, C.; Agarwal, A.; Ding, T.; Tien, H.-J.; Akey, A.; Gardener, J.; Singh, B.; Watanabe, K.; Taniguchi, T.; Burch, K. S.; Bell, D. C.; Zhou, B. B.; Gao, W.; Lu, H.-Z.; Bansil, A.; Lin, H.; Chang, T.-R.; Fu, L.; Ma, Q.; Ni, N.; Xu, S.-Y. Layer Hall Effect in a 2D Topological Axion Antiferromagnet. *Nature* **2021**, *595* (7868), 521–525.

(10) Du, L.; Hasan, T.; Castellanos-Gomez, A.; Liu, G.-B.; Yao, Y.; Lau, C. N.; Sun, Z. Engineering Symmetry Breaking in 2D Layered Materials. *Nat. Rev. Phys.* **2021**, *3* (3), 193–206.

(11) Aivazian, G.; Gong, Z.; Jones, A. M.; Chu, R.-L.; Yan, J.; Mandrus, D. G.; Zhang, C.; Cobden, D.; Yao, W.; Xu, X. Magnetic Control of Valley Pseudospin in Monolayer WSe$_2$. *Nat. Phys.* **2015**, *11* (2), 148–152.

(12) Sun, Z.; Yi, Y.; Song, T.; Clark, G.; Huang, B.; Shan, Y.; Wu, S.; Huang, D.; Gao, C.; Chen, Z.; McGuire, M.; Cao, T.; Xiao, D.; Liu, W.-T.; Yao,

W.; Xu, X.; Wu, S. Giant Nonreciprocal Second-Harmonic Generation from Antiferromagnetic Bilayer CrI$_3$. *Nature* **2019**, *572* (7770), 497–501.

(13) Song, T.; Cai, X.; Tu, M. W.-Y.; Zhang, X.; Huang, B.; Wilson, N. P.; Seyler, K. L.; Zhu, L.; Taniguchi, T.; Watanabe, K.; McGuire, M. A.; Cobden, D. H.; Xiao, D.; Yao, W.; Xu, X. Giant Tunneling Magnetoresistance in Spin-Filter van Der Waals Heterostructures. *Science* **2018**, *360* (6394), 1214–1218.

(14) Huang, B.; Clark, G.; Navarro-Moratalla, E.; Klein, D. R.; Cheng, R.; Seyler, K. L.; Zhong, D.; Schmidgall, E.; McGuire, M. A.; Cobden, D. H.; Yao, W.; Xiao, D.; Jarillo-Herrero, P.; Xu, X. Layer-Dependent Ferromagnetism in a van Der Waals Crystal down to the Monolayer Limit. *Nature* **2017**, *546* (7657), 270–273.

(15) Telford, E. J.; Dismukes, A. H.; Lee, K.; Cheng, M.; Wieteska, A.; Bartholomew, A. K.; Chen, Y.-S.; Xu, X.; Pasupathy, A. N.; Zhu, X.; Dean, C. R.; Roy, X. Layered Antiferromagnetism Induces Large Negative Magnetoresistance in the van Der Waals Semiconductor CrSBr. *Adv. Mater.* **2020**, *32* (37), 2003240.

(16) Lee, K.; Dismukes, A. H.; Telford, E. J.; Wiscons, R. A.; Wang, J.; Xu, X.; Nuckolls, C.; Dean, C. R.; Roy, X.; Zhu, X. Magnetic Order and Symmetry in the 2D Semiconductor CrSBr. *Nano Lett.* **2021**, *21* (8), 3511–3517.

(17) Ghiasi, T. S.; Kaverzin, A. A.; Dismukes, A. H.; de Wal, D. K.; Roy, X.; van Wees, B. J. Electrical and Thermal Generation of Spin Currents by Magnetic Bilayer Graphene. *Nat. Nanotechnol.* **2021**, *16* (7), 788–794.

(18) Wilson, N. P.; Lee, K.; Cenker, J.; Xie, K.; Dismukes, A. H.; Telford, E. J.; Fonseca, J.; Sivakumar, S.; Dean, C.; Cao, T.; Roy, X.; Xu, X.; Zhu, X. Interlayer Electronic Coupling on Demand in a 2D Magnetic Semiconductor. *Nat. Mater.* **2021**.

(19) Wang, H.; Qi, J.; Qian, X. Electrically Tunable High Curie Temperature Two-Dimensional Ferromagnetism in van Der Waals Layered Crystals.

*Appl. Phys. Lett.* **2020**, *117* (8), 83102.

(20) Wang, C.; Zhou, X.; Zhou, L.; Tong, N.-H.; Lu, Z.-Y.; Ji, W. A Family of High-Temperature Ferromagnetic Monolayers with Locked Spin-Dichroism-Mobility Anisotropy: MnNX and CrCX (X = Cl, Br, I; C = S, Se, Te). *Sci. Bull.* **2019**, *64* (5), 293–300.

(21) Gong, C.; Li, L.; Li, Z.; Ji, H.; Stern, A.; Xia, Y.; Cao, T.; Bao, W.; Wang, C.; Wang, Y.; Qiu, Z. Q.; Cava, R. J.; Louie, S. G.; Xia, J.; Zhang, X. Discovery of Intrinsic Ferromagnetism in Two-Dimensional van Der Waals Crystals. *Nature* **2017**, *546* (7657), 265–269.

(22) Huang, B.; McGuire, M. A.; May, A. F.; Xiao, D.; Jarillo-Herrero, P.; Xu, X. Emergent Phenomena and Proximity Effects in Two-Dimensional Magnets and Heterostructures. *Nat. Mater.* **2020**, *19* (12), 1276–1289.

(23) Li, B.; Wan, Z.; Wang, C.; Chen, P.; Huang, B.; Cheng, X.; Qian, Q.; Li, J.; Zhang, Z.; Sun, G.; Zhao, B.; Ma, H.; Wu, R.; Wei, Z.; Liu, Y.; Liao, L.; Ye, Y.; Huang, Y.; Xu, X.; Duan, X.; Ji, W.; Duan, X. Van Der Waals Epitaxial Growth of Air-Stable CrSe$_2$ Nanosheets with Thickness-Tunable Magnetic Order. *Nat. Mater.* **2021**, *20* (6), 818–825.

(24) Jiang, Z.; Wang, P.; Xing, J.; Jiang, X.; Zhao, J. Screening and Design of Novel 2D Ferromagnetic Materials with High Curie Temperature above Room Temperature. *ACS Appl. Mater. Interfaces* **2018**, *10* (45), 39032–39039.

(25) Bonilla, M.; Kolekar, S.; Ma, Y.; Diaz, H. C.; Kalappattil, V.; Das, R.; Eggers, T.; Gutierrez, H. R.; Phan, M.-H.; Batzill, M. Strong Room-Temperature Ferromagnetism in VSe$_2$ Monolayers on van Der Waals Substrates. *Nat. Nanotechnol.* **2018**, *13* (4), 289–293.

(26) Williams, T. J.; Aczel, A. A.; Lumsden, M. D.; Nagler, S. E.; Stone, M. B.; Yan, J.-Q.; Mandrus, D. Magnetic Correlations in the Quasi-Two-Dimensional Semiconducting Ferromagnet CrSiTe$_3$. *Phys. Rev. B* **2015**, *92* (14), 144404.

(27) Worledge, D. C.; Geballe, T. H. Magnetoresistive Double Spin Filter

Tunnel Junction. *J. Appl. Phys.* **2000**, *88* (9), 5277–5279.

(28) Klein, D. R.; MacNeill, D.; Lado, J. L.; Soriano, D.; Navarro-Moratalla, E.; Watanabe, K.; Taniguchi, T.; Manni, S.; Canfield, P.; Fernández-Rossier, J.; Jarillo-Herrero, P. Probing Magnetism in 2D van Der Waals Crystalline Insulators via Electron Tunneling. *Science* **2018**, *360* (6394), 1218–1222.

(29) Baibich, M. N.; Broto, J. M.; Fert, A.; Van Dau, F. N.; Petroff, F.; Etienne, P.; Creuzet, G.; Friederich, A.; Chazelas, J. Giant Magnetoresistance of (001)Fe/(001)Cr Magnetic Superlattices. *Phys. Rev. Lett.* **1988**, *61* (21), 2472–2475.

(30) Telford, E. J.; Dismukes, A. H.; Dudley, R. L.; Wiscons, R. A.; Lee, K.; Chica, D. G.; Ziebel, M. E.; Han, M.-G.; Yu, J.; Shabani, S.; Scheie, A.; Watanabe, K.; Taniguchi, T.; Xiao, D.; Zhu, Y.; Pasupathy, A. N.; Nuckolls, C.; Zhu, X.; Dean, C. R.; Roy, X. Coupling between Magnetic Order and Charge Transport in a Two-Dimensional Magnetic Semiconductor. *Nat. Mater.* **2022**.

(31) Yang, S.; Xu, X.; Zhu, Y.; Niu, R.; Xu, C.; Peng, Y.; Cheng, X.; Jia, X.; Huang, Y.; Xu, X.; Lu, J.; Ye, Y. Odd-Even Layer-Number Effect and Layer-Dependent Magnetic Phase Diagrams in $MnBi_2Te_4$. *Phys. Rev. X* **2021**, *11* (1), 11003.

(32) Lyons, T. P.; Gillard, D.; Molina-Sánchez, A.; Misra, A.; Withers, F.; Keatley, P. S.; Kozikov, A.; Taniguchi, T.; Watanabe, K.; Novoselov, K. S.; Fernández-Rossier, J.; Tartakovskii, A. I. Interplay between Spin Proximity Effect and Charge-Dependent Exciton Dynamics in $MoSe_2$/$CrBr_3$ van Der Waals Heterostructures. *Nat. Commun.* **2020**, *11* (1), 6021.

(33) Yang, K.; Wang, G.; Liu, L.; Lu, D.; Wu, H. Triaxial Magnetic Anisotropy in the Two-Dimensional Ferromagnetic Semiconductor CrSBr. *Phys. Rev. B* **2021**, *104* (14), 144416.

(34) Huang, B.; Clark, G.; Klein, D. R.; MacNeill, D.; Navarro-Moratalla, E.;

Seyler, K. L.; Wilson, N.; McGuire, M. A.; Cobden, D. H.; Xiao, D.; Yao, W.; Jarillo-Herrero, P.; Xu, X. Electrical Control of 2D Magnetism in Bilayer CrI$_3$. *Nat. Nanotechnol.* **2018**, *13* (7), 544–548.

(35) Wang, Z.; Gutiérrez-Lezama, I.; Ubrig, N.; Kroner, M.; Gibertini, M.; Taniguchi, T.; Watanabe, K.; Imamoğlu, A.; Giannini, E.; Morpurgo, A. F. Very Large Tunneling Magnetoresistance in Layered Magnetic Semiconductor CrI$_3$. *Nat. Commun.* **2018**, *9* (1), 2516.

**For Table of Contents Only**

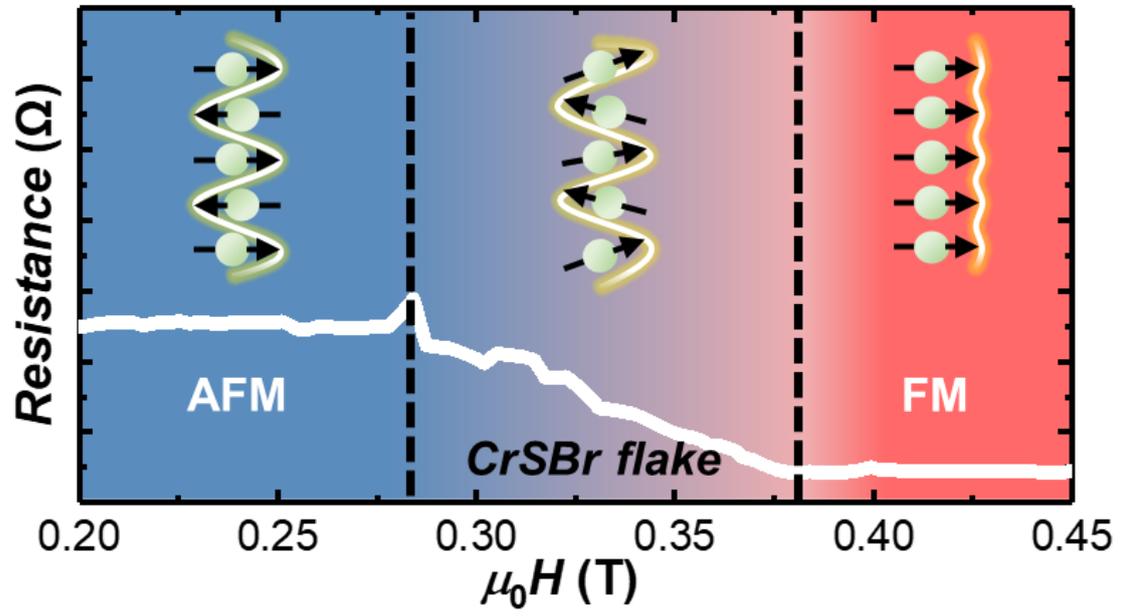